\documentclass[12pt]{article}
\usepackage{epsfig}
\usepackage{color}

\def\beq{\begin{equation}}
\def\eeq#1{\label{#1}\end{equation}}
\def\eeqn{\end{equation}}

\def\beqa{\begin{eqnarray}}
\def\eeqa#1{\label{#1}\end{eqnarray}}
\def\eeqan{\end{eqnarray}}

\let\bar=\overbar

\def\Dslash{\not{\hbox{\kern-4pt $D$}}}
\def\dslash{\not{\hbox{\kern-2pt $\del$}}}

\def\msb{{\bar{\ssstyle M \kern -1pt S}}}

\usepackage{fancyhdr,graphicx}
\def\Title#1{\begin{center} {\Large {\bf #1} } \end{center}}

\begin{document}

\Title{Are AXPs/SGRs magnetars?}

\bigskip\bigskip


\begin{raggedright}

{\it \begin{center} { Qiao G. J.$^{1}$, Xu R. X.$^{1}$, and 
Du Y. J.$^{2}$ }\end{center} 

$^{1}$Department of Astronomy, Peking University, Beijing 100871,
China

$^{2}$National Astronomical Observatories, Chinese Academy of
Sciences, Jia-20, Datun Road, Chaoyang District, Beijing 100012, China \\
 }
\bigskip\bigskip
\end{raggedright}

\begin{abstract}

Anomalous X-ray Pulsars and Soft Gamma-Ray Repeaters have been
generally recognized as neutron stars with super strong magnetic
fields, namely ``magnetars". The ``magnetars" manifest that the
luminosity in X-ray band are larger than the rotational energy loss
rate, i.e. $L_{X}>\dot {E}_{\rm rot}$, and then the radiation energy
is coming from the energy of magnetic field. Here it is argued that
magnetars may not really exist. Some X-ray and radio observational
results are contradicted with the magnetar model.

(1) The X-ray luminosity of PSR J1852+0040 is much larger than the
rotational energy loss rate ($L_{X}/\dot {E}_{\rm rot}\simeq 18 )$,
but the magnetic field is just $3.1\times 10^{11}$\,G. Does this X-ray
radiation energy come from the magnetic field?

(2) In contrast to the above, the magnetic fields of radio pulsars
J1847-0130 and PSR J1718-3718 are higher than that of AXP 1E 2259+586,
why is the radiation energy of those two radio pulsars still coming
from rotational energy? Furthermore, the magnetic field
  of the newly discovered SGR 0418+5729 with the lowest magnetic field
  is 3.0\,e13\,G, lower than the critical magnetic field $B_{\rm
    C}=4.414$\,e13\,G) (Esposito et al. 2010).

(3) Some ``magnetars" also emit normal transient radio pulses, what is
the essential difference between radio pulsars and the ``magnetars"?

The observational fact arguments will be presented at first, then we
discuss in what situation the conventional method to obtain magnetic
field could not be correct.

\end{abstract}

\section{Introduction}

Anomalous X-ray Pulsars (AXPs) and Soft Gamma-Ray Repeaters (SGRs)
have similar characteristics (see, e.g. Mereghetti et al., 2009;
Kaspi, 2007; 2009). Timing analysis gives no evidence of orbit motion,
no evidence to show AXPs and SGRs in binary systems. Hulleman et
al. (2000) found that the optical emission from the AXP (4U 0124+61) is
too faint to admit a large accretion disk.  AXPs and SGRs are
recognized as the new classes of young objects ($P/(2 \dot P) \sim
10^3 - 10^5 $\,yr), along with other young neutron star families, and
the magnetic fields are super strong ($5.9\times10^{13}$\,G to
$10^{15}$\,G). They do not manifest themselves as radio pulsars, like
the dim isolated neutron stars (DINs) and the center compact objects
(CCOs), which suggests alternative evolutionary paths for young
neutron stars (Ertan et al. 2009).  The most peculiar observed fact is
that all AXPs and SGRs have stable persistent pulsed X-ray emission
with the luminosity of $L_{X} \sim 10^{34} - 10^{36}$\,ergs\,s$^{-1}$,
well in excess of the spin down energy of these sources, i.e.,
$L_{X}>\dot{E}_{\rm rot}$.  A very important question is then: what
does the radiation energy come from?  To answer this question, some
models are suggested. Meanwhile, the ``magnetar" model is widely
accepted, which is isolated neutron stars with super strong magnetic
fields (Duncan \& Thompson 1992; Thompson \& Doncan 1995), and the
radiation energy is coming from the magnetic field decay (Thompson \&
Doncan 1996).

For AXPs and SGRs, the widely accepted assumption is that the
rotational energy is carried out by the dipole magnetic radiation.
This is just the fact what we have observed for radio pulsars. So
radio pulsars are named as ``spin-down powered" neutron
stars. Unfortunately for AXPs and SGRs, no direct link between
rotational energy and the dipole magnetic radiation has been found so
far, thus any calculations for the value of magnetic field obtained
from $P$ and $\dot{P}$ could not be reliable.

Furthermore, there are some observations that challenge the
``magnetar" model.

(1) X-ray observations show at least one CCO, PSR J1852+0400, in the
supernova remnants Kes 79, shows ``anti-magnetar" properties: the
field is low ($3.1\times 10^{10}$\,G), but X-ray emission is larger
than the spin down energy, $L_{X}>\dot{E}_{\rm rot}$. What does the
radiation energy come from? Does it also come from the magnetic field?

(2) Some authors argue that owing to the super strong magnetic field,
so AXPs and SGRs are different from the normal radio pulsars:
magnetars can not radiate in the radio band; and radio pulsars can not
have super strong magnetic field. But recent observations show that
some AXPs: XTE J1810-197 (Camilo et al. 2007) and 1E 1547.0-5408
(Camilo et al. 2008) can have normal radio pulse emission. On the other
hand, some normal radio pulsars PSR J1847-0130 and PSR J1718-3718 do
have super strong magnetic field , which are larger than that of one
of the AXPs. In this case, can we say that the differences between
``magnetars" and radio pulsars are coming from the difference of the
magnetic field? If not, what is the real difference?

In $\S$ 2, we present the anti-magnetar observations in both X-ray and
radio bands.  In $\S$3, it is argued that the magnetic field values of
``magnetsrs" are incorrect.  The conclusion and discussion are
presented in $\S 4$.

\section{Anti-magnetar observations: challenge to magnetars}

Some anti-magnetar observations in both X-ray and radio bands are
presented below, which challenges the existence of the magnetars with
super-strong magnetic field.

\subsection{X-ray observations: $L_{x}>\dot{E}_{rot}$ does not mean
super-strong magnetic field at all}

Halpern \& Gotthelf (2010), using the data of XMM-Newton and Chandra,
achieved phase-connected timing of the 105\,ms X-ray pulsar PSR
J1852+0040 that provides the first measurement of the spin-down rate
of a CCO in a supernova remnants (See table 1). Some observations
challenge the ``magnetars":

\begin{table}[] \begin{center}
\caption{The parameters: spin period ($P$), period derivative
  ($\dot{P}$), characteristic age ($\tau_{c}$) and Host object age
  ($\tau_{\rm host}$) and the Host name for three CCOs.
 \label{tab:publ-works}} \begin{tabular}{|l|c|c|c|c|c|}
\hline
Name & $P$\,(ms) & $\dot{P}$\,(s$\rm \cdot s^{-1}$) &
$\tau_{c}$\,(yr) & $\tau_{host}$(yr) & Host \\
%
\hline
PSR J1852+0040 & 105 & 8.7e-18 & 192e6 & $\sim 7$e3 &
  Kes 75 (a) \\
\hline 1E 1207.4.5209& 424 & 6.6e-17 & $>$27e6 & $\sim 7$e3 & SNR PKS
1209.51/52 (b) \\
\hline
RX J0822-4300 & 112 & $<$8.3e-15 & $>$0.22e6 & 3.7e3  &  SNR Puppis A (c) \\
\hline\hline

    \end{tabular} \end{center}
(a)A CCO, in Kes 75 (Halpern \& Gotthelf 2010).\\
(b)A CCO, in SNR PKS 1209.51/52 (Gotthelf and Halpern 2008).\\
(c)A CCO, in SNR Puppis A (Halpern \& Gotthelf 2010).

\end{table}

(1) $L_{X}>\dot{E}_{\rm rot}$ but the surface magnetic strength
$B_{s}$ is low.

PSR J1852+0040, $P=105$ ms, $\dot{P} = (8.68\pm 0.09)\times
10^{-18}~\rm \,s\,s^{-1}$, the surface magnetic field is $B_{s} =
3.1\times10^{10}$\,G, which is the weakest magnetic field ever
measured for a young neutron star. The X-ray luminosity $L_{X} =
5.3\times 10^{33}(d/7.1$ kpc$)^2$\,erg\,s$^{-1}$.  The rotational
energy loss is $\dot{E}_{\rm rot} = 3.0 \times 10^{32}$\,erg\,s$^{-1}$,
then $L_{X}/\dot{E}_{\rm rot}\simeq 17.7$.

Is the radiation energy still coming from the magnetic field?

\begin{table}[]
\begin{center}
\caption{A comparison between radio pulsars, AXPs and an
  anti-magnetsr.
%
 \label{tab:publ-works}}

\begin{tabular}{|l|c|c|c|c|c|}
\hline
Name & $P$\,(s) & $\dot{P}$\,(s$\rm \cdot s^{-1}$) &
$\tau_{c}$\,(yr) & $B_{\rm s}$\,(G) & Note \\
\hline
PSR J1847-0130 & 6.7 & 1.3e-12 & 8.2e4 & 9.4e13 & High-B Radio PSR(a)  \\
\hline
PSR J1718-3718 & 3.3 & 1.5e-12 & 3.5e4 & 7.4e13  &
  High-B Radio PSR(b) \\
\hline
PSR J1814-1733 & 4.0 & 7.4e-13 & 8.6e4 & 5.5e13  & High B Radio PSR(c) \\
\hline
\hline
\hline
SGR 0418+5729 & 9.1 & 1.1e-13 & 4.1e6 & 3.0e13 &
Magnetar, no radio (d)\\
\hline
1E2255+586& 7.0 & 4.9e-13 & 2.3e5 & 5.9e13 &
Magnetar, no radio(e)\\
\hline
XTEJ1810-197& 2.1 & 2.3e-11 & 1.4e3 & 2.4e14 &
Magnetar, radio(f)\\
\hline
1E 1547.0-5408& 5.54 & 1.2e-11 & 7.6e3 & 2.6e14 &
Magnetar, radio(g)\\
\hline
\hline
\hline
PSR J1846-0258 & 0.324 & 7.1e-12 & 7.2e2 & 4.9e13 &
 $L_{X}/\dot{E}_{\rm
  rot}=0.05$,no radio(h)\\
\hline
PSR J1852+0040 & 0.105 & 8.7e-18 & 1.9e8 & 3.1e10
& Anti-mag.$L_{X}/\dot{E}=17.7$,no radio(i)\\
%
\hline

    \end{tabular} \end{center}
(a) High magnetic field pulsar (High-B PSR) (Mclaughlin et al. 2003);
no X-ray.\\
(b) High-B PSR (Hobbs, G., et al. 2004; Kaspi1 \& McLaughlin 2005);
low X-ray.\\
(c) High-B PSR (Camilo et al. 2000); no X-ray. \\
(d) A newly discovered SGR with the lowest magnetic field (lower than
the critical magnetic field $B_{\rm C}=4.414e13$\,G) (Esposito et al. 2010).\\
(e) A glitching AXP (Kaspi 2007). \\
(f) A transient radio-emitting AXP (Halpern et al. 2005; Camilo et
al. 2006).  \\
(g) A transient radio-emitting AXP (Camilo et al. 2007). \\
(h) A rotation-powered pulsar with High-B, $L_{X}/\dot{E}_{\rm
  rot}=0.05$; surrounded by a PWN and a SNR shell in Kes 75; no radio
emission.  There are AXP-like bursts (Gotthelf et al. 2000; Archibald
et al. 2008). \\
(i) A CCO (``anti-magnetar") in the center of the SNR Kes75,
$L_{X}/\dot{E}=17.7$; no radio (Halpern \& Gotthelf 2010).

\end{table}

(2) The characteristic ages of CCOs are not consistent with the companied
SNR.

All CCOs are in SNRs (Halpern \& Gotthelf 2010). The characteristic
ages, $\tau_{c}=P/(2 \dot{P})$, are not consistent with the ages of
companied SNRs.  For example, the characteristic age $\tau_{c}$ of PSR
J1852+0040 is 192\,Myr, but the age of companied SNR Kes 79 is 7\,kyr.

This problem also appears for other CCOs. For example, the
characteristic age of the CCO RX J0822-4300 (P=112 ms) is $\tau_{c}>
220$\,kyr, but the age of companied SNR Puppis A is 3.\, kyr (Gotthelf
\& Halpern 2009).  The characteristic age of the central source 1E
1207.4-5209 (p=424 ms) in the SNR PKS 1209.51/52 is $\tau_{c}>
27$\,Myr or $\tau_{c}=200-900$ kyr (Shi and Xu 2003), but the age of
companied SNR PKS 1209.51/52 is 7\,kyr (Gotthelf \& Halpern 2008).

Where does the problem come from? Is the calculated characteristic age
correct? If the characteristic age calculated from $P$ and $\dot P$ is
incorrect, how about the magnetic field?

\subsection{Radio observations: the difference between radio pulsars
and magnetars does not originate from the difference of magnetic fields}


 Previous observations mainly show that: (1) no radio emission from
 ``magnetars" are observed; (2) the magnetic fields of ``magnetars"
 are stronger than those of normal radio pulsars.  It is then
 generally believed that these differences are caused by the
 difference of the magnetic fields.

Recent observations show that all these two differences are confused:
some radio pulsars have stronger magnetic field than that of one AXP;
radio emission from some AXPs are observed clearly after X-ray flare.

\subsubsection{The magnetic field of radio pulsars larger than
one  ``magnetsrs"}

In table 2, a comparison between four high-B radio pulsars and three
AXPs is presented. One can see that the magnetic fields of PSR
J1847-0130 ($9.4 \times 10^{13}$\,G) and PSR J1718-3718 ($7.4 \times
10^{13}$\,G) are lager than that of 1E 2255+586 ($5.9\times
10^{13}$\,G). Beside this, the magnetic fields of few other pulsars are
closer to that of 1E 2255+586. Such as PSR J1814-1733, the magnetic
field is $5.5 \times 10^{13}$\,G.

PSR J1846-0258 is a rotation-powered pulsar with High-B. The magnetic
field is $4.9\times10^{13}$\,G.  The X-ray emission luminosity is
smaller than that of the spin-down energy, e.g.  $L_{X}/\dot{E}_{\rm
  rot}=0.05$.  It is surrounded by a PWN and a SNR shell, in Kes
75. This pulsar has not been observed in radio band, but there are
AXP-like bursts and magnetar like transition (Gotthelf et al. 2000;
Archibald et al. 2008).

All these observations present a confusion, do the differences of
these objects come the difference of the magnetic field strength?

\subsubsection{Radio emission observed from some ``magnetsrs"}

A typical radio emission has been observed from AXP XTE J1810-197 in
2004 January, one year after outburst (Halpern et al. 2005; Camilo et
al. 2006; Lazaridis et al. 2008).  The period of XTE J1810-197 is
$P=5.54$ s, the magnetic field is $2.6 \times 10^{14}$\,G.  The radio
  pulse profiles are observed at many frequencies (Camilo et
  al. 2006).

Another AXP 1E 1547.0-5408 also observed transient pulsed radio
emission following X-ray bursts (Camilo et al. 2007). The period of 1E
1547.0-5408 is $P=2.069$\,s, with a surface field strength of
$B_{s}=2.2\times10^{14} $\,G.

The pulsed radio emission of these two objects are similar as the
typical emission from normal radio pulsars.  Some authors argue that
the radio spectrum of AXPs is approximately flat, which is different
from normal radio pulsars. But there are 10 percent of radio pulsars
have flat spectrum (Dick Manchester, private conversation). After the
pulsed radio emissions from AXPs have been observed, some authors took
calculations to show that, in the case of strong magnetic field, the
radio emission could still be observed. Our question is
  that if this is the case, why was the pulsed radio emission radiated
  just after outburst and observed only in some time?


\section{Do super strong magnetic fields really exist in AXPs \& SGRs?  }

As we discussed above, all characteristic ages of CCOs do not agree
with the ages of the related SNRs.  If we believe that the ages of
SNRs are correct in most situations, then we should ask why the
characteristic ages are different from each other so much? Here it is
argued that, the values of magnetic field, rotational energy loss rate
and characteristic age obtained from $P$ and $\dot P$ are incorrect
for AXPs and SGRs.

How about the magnetars? From the Table 3, one can see that the same
thing happens in the case of ``magnetars".  The characteristic ages
($\tau_{c}$) of ``magnetars" and the Host object ages ($\tau_{\rm
  host}$) are very different (e.g. Shi \& Xu 2003).

Our suggestion is that: using period $P$ and period derivative
$\dot{P}$ to get the super strong magnetic fields is incorrect for
AXPs and SGRs!

\begin{table}[] \begin{center}
\caption{The characteristic ages ($\tau_{c}$) and the Host object
  ages ($\tau_{\rm host}$) of some magnetars.
 \label{tab:publ-works}} \begin{tabular}{|l|c|c|c|c|c|}
\hline
Name & $P$\,(s) & $\dot{P}$\,(s$\rm \cdot s^{-1}$) &
$\tau_{c}$\,(yr) & $\tau_{host}$(yr) & Host \\
\hline
SGR 1806-20 & 7.46 & 10e-10 & 1.2e3 & (3-5)e6 & MSC (a)  \\
\hline
SGR 1900+14 & 5.16 & 10e-10 & 0.8e3 & (1-10)e6  &
 SNR: CTB 109 (b) \\
\hline
1E 2259+586 & 7.0 & 4.84e-13 & 2.28e5 & 1.7e4  & SNR: CTB 109 (c) \\
%
\hline\hline

    \end{tabular} \end{center}
(a) A magnetar in the Host object of massive star cluster (MSC) (Fuchs et
  al. 1999; Bibby et al. 2000). \\
(b) A magnetar in the Host object of CTB 109 (Vrba et al. 2000). \\
(c) A magnetar in the Host object of CTB 109 (Hughes et al. 1981;
  Iwasawa et al. 1992; Marsden et al. 2001).
\end{table}

\section{Conclusion \& discussion}

From the discussions above, our conclusions are drawn as follows:

1. An anti-magnetar (PSR J1852+0040) shows that the X-ray luminosity
is larger than the rotational energy loss rate, that is
$L_{X}/\dot{E}>1$. The magnetic field is $3.1\times
10^{10}$\,G. Where does the radiation energy come from? Can we say that it
is also coming from the magnetic energy?

We argue that in the case of $L_{X}/\dot{E}>1$, using $P$ and
$\dot{P}$ to get the magnetic field is incorrect, flow-out particles
and the parallel magnetic component, $\mu_{\parallel}$, should be
taken into account, which contributes to the rotational energy loss
(Xu \& Qiao 2001).

2. The differences of ``magnetars" and the radio pulsars are not from
the difference of the magnetic field strength. The transient pulsed
radio emission from ``magnetars" is a kind of typical radio emission
as similar as the one observed for normal radio pulsars. The flat
spectrum of magnetars is not different from normal radio pulsars. An
important thing is that, to show the distinct difference between the
transient pulsed radio emission from the magnetars (that is related to
the outburst) and the normal radio pulsars.

3. Many authors show that how to produce the observed phenomena of
magnetars, but never to show the differences between normal radio
pulsar and magnetars. For example, why magnetars have high energy
radiations but nor for the radio pulsar PSR J1847-0130.

We argue here that a acceptable theory, related to observations of
manetars, should show the difference between magnetars and other
objects.

4. Producing the very strong outbursts is a key question for any model
about the AXPs \& SGRs. If the very high energy of the outburst does
not come from the magnetic field, where oes the energy come from? Only
accretion may not satisfy this request. Some possible quark star
models should be taken into account, e.g. Cheng \& Dai (1998), Cheng et
al. (2000); Xu, Tao \& Yang (2006).



\def\Discussion{
\setlength{\parskip}{0.3cm}\setlength{\parindent}{0.0cm}
     \bigskip\bigskip      {\Large {\bf Discussion}} \bigskip}
\def\speaker#1{{\bf #1:}\ }
\def\endDiscussion{}

\end{document}